\documentclass[sigconf]{acmart}

\usepackage{threeparttable}
\usepackage{multirow}
\usepackage{makecell}
\usepackage{colortbl}
\usepackage{amsthm,amsmath}
\usepackage{marvosym}
\usepackage{mathrsfs}

\usepackage{subcaption}
\usepackage{graphicx}
\AtBeginDocument{%
  \providecommand\BibTeX{{%
    \normalfont B\kern-0.5em{\scshape i\kern-0.25em b}\kern-0.8em\TeX}}}

\setcopyright{acmcopyright}
\copyrightyear{2023}
\acmYear{2023}
\acmDOI{XXXXXXX.XXXXXXX}

\acmConference[SIGIR '24]{The 47th International ACM SIGIR Conference on Research and Development in Information Retrieval}{July 14-18, 2024}{Washington D.C., USA}
\acmPrice{15.00}
\acmISBN{978-1-4503-XXXX-X/18/06}

\begin{document}

\title[EEG-SVRec]{EEG-SVRec: An EEG Dataset with User Multidimensional Affective Engagement Labels in Short Video Recommendation}


\author{Shaorun Zhang}
\authornote{These authors contributed equally to this work.\label{first-co}}
\email{zhangshaorun@126.com}
\orcid{0009-0006-3287-2956}

\affiliation{%
  \institution{DCST, Tsinghua University}
  \institution{Quan Cheng Laboratory}
  \institution{Zhongguancun Laboratory}
  \city{Beijing}
  \country{China}
}

\author{Zhiyu He\textsuperscript{\ref{first-co}}}
\orcid{0000-0003-1291-2739}
\affiliation{%
  \institution{DCST, Tsinghua University}
  \institution{Quan Cheng Laboratory}
  \institution{Zhongguancun Laboratory}
  \city{Beijing}
  \country{China}
}

\author{Ziyi Ye}
\orcid{0000-0002-5622-0235}
\affiliation{%
  \institution{DCST, Tsinghua University}
  \institution{Quan Cheng Laboratory}
  \institution{Zhongguancun Laboratory}
  \city{Beijing}
  \country{China}
}

\author{Peijie Sun}
\orcid{0000-0003-1291-2739}
\affiliation{%
  \institution{DCST, Tsinghua University}
  \institution{Quan Cheng Laboratory}
  \institution{Zhongguancun Laboratory}
  \city{Beijing}
  \country{China}
}

\author{Qingyao Ai} 
\authornote{Corresponding Authors.\label{correspond}}
\orcid{0000-0002-5030-709X}
\affiliation{%
  \institution{DCST, Tsinghua University}
  \institution{Quan Cheng Laboratory}
  \institution{Zhongguancun Laboratory}
  \city{Beijing}
  \country{China}
}

\author{Min Zhang\textsuperscript{\ref{correspond}}}
\orcid{0000-0003-3158-1920}
\affiliation{%
  \institution{DCST, Tsinghua University}
  \institution{Quan Cheng Laboratory}
  \institution{Zhongguancun Laboratory}
  \city{Beijing}
  \country{China}
}

\author{Yiqun Liu}
\orcid{0000-0002-0140-4512}
\affiliation{%
  \institution{DCST, Tsinghua University}
  \institution{Quan Cheng Laboratory}
  \institution{Zhongguancun Laboratory}
  \city{Beijing}
  \country{China}
}
\renewcommand{\shortauthors}{Zhang and He, et al.}
\begin{abstract}
In recent years, short video platforms have gained widespread popularity, making the quality of video recommendations crucial for retaining users. Existing recommendation systems primarily rely on behavioral data, which faces limitations when inferring user preferences due to issues such as data sparsity and noise from accidental interactions or personal habits. To address these challenges and provide a more comprehensive understanding of user affective experience and cognitive activity, we propose EEG-SVRec, the first \textbf{EEG} dataset with User Multidimensional Affective Engagement Labels in \textbf{S}hort \textbf{V}ideo \textbf{Rec}ommendation.

The study involves 30 participants and collects 3,657 interactions, offering a rich dataset that can be used for a deeper exploration of user preference and cognitive activity. By incorporating self-assessment techniques and real-time, low-cost EEG signals, we offer a more detailed understanding user affective experiences~(valence, arousal, immersion, interest, visual and auditory) and the cognitive mechanisms behind their behavior.
We establish benchmarks for rating prediction by the recommendation algorithm, showing significant improvement with the inclusion of EEG signals. Furthermore, we demonstrate the potential of this dataset in gaining insights into the affective experience and cognitive activity behind user behaviors in recommender systems. This work presents a novel perspective for enhancing short video recommendation by leveraging the rich information contained in EEG signals and multidimensional affective engagement scores, paving the way for future research in short video recommendation systems.

\end{abstract}



\maketitle

\section{Introduction}
\label{intro}


In recent years, short videos have emerged as a popular medium for entertainment and communication across various social media platforms, attracting millions of users worldwide. These videos typically span from a few seconds to several minutes and encompass a broad spectrum of content. 
Short video platforms generally gather, process, and analyze user behavior data and video information. To enhance the recommendation quality and retain users, various recommendation strategies are employed, including interest-based recommendations~\cite{gou2011sfviz, ye2011exploiting}, popularity-based recommendations~\cite{bressan2016limits, yang2016effects}, and personalized recommendations~\cite{zhang2021commentary}. The choice of a recommendation strategy can significantly impact users' affective engagement while browsing short videos. 

Existing short video recommendation systems mainly focus on behavioral metrics, such as likes, dwell time, view percentage, etc., to improve recommendation performance~\cite{jannach2018recommending, shani2011evaluating,sun2023neighborhood}. These behavior data are usually collected from user logs and applied as implicit feedback signals to infer user preferences.
Although these observed data usually contain abundant information, only considering existing information is not enough to gain a comprehensive understanding of users~\cite{lu2019quality, baeza2018bias}. There still exist challenges in capturing user preference from behavioral data. Firstly, behavioral data, such as likes and comments, is usually sparse. Secondly, the presence of noise, resulting from accidental interactions or personal habits, can affect the reliability of the data.

In order to deeply understand users' cognitive activities, we record EEG~(Electroencephalograph) signals during short video browsing. EEG, as a neuroelectrical signal, containing rich spatial, temporal, and frequency band information about human experience, can be used to study the underlying neural mechanisms and can reflect relevant information about user cognition, emotion, and attention~\cite{teplan2002fundamentals,li2022eeg, moshfeghi2016understanding, ye2022towards}. Providing high temporal resolution data, the application of EEG technology in the Information Retrieval~(IR) domain has been proven to be useful~\cite{ye2022don, davis2021collaborative}. 
At the same time, the latest developments in EEG recording devices are known for their high portability and low operating costs~\cite{hu2019ten, rashid2020current}, which are necessary for real-world application scenarios. The high temporal resolution of EEG data enables it to effectively address the real-time demands of short video recommendation scenarios.

To further understand the relationship between user behavior and EEG signals, it is essential to incorporate user affective experiences into the annotation of short videos. These affective experiences are from different dimensions. Emotion elicited by short videos plays a significant role in the browsing experience, which is commonly modeled via two dimensions: valence and arousal~\cite{thayer1990biopsychology}. 
Both the degree to which short videos align with user interest and the level of immersion experienced by users while browsing short videos influence user behavior and perception.
Besides, short videos serve as a combined visual and auditory medium, so understanding the impact of visual and auditory features on users' perceptions can be helpful. 
Accordingly, we collect six Multidimensional Affective Engagement Scores~(MAES), which are valence, arousal, immersion, interest, visual and auditory, and extract the visual and auditory features of the videos.

By employing self-assessment techniques, we obtain a more detailed and multidimensional perspective of user experience. Furthermore, real-time, low-cost EEG signals can be utilized to gain insights into users' cognitive activity.




Therefore, We proposed to build EEG-VSRec\footnote{Dataset and codes are available at 
https://anonymous.4open.science/r/Z-SV-CFB1\label{link}}, an \textbf{E}EG dataset with user Multidimensional Affective Engagement in \textbf{S}hort \textbf{V}ideo \textbf{Rec}ommendations. We conducted the user study where participants continuously viewed short videos in several sessions. After each session, the participants rated the MAES for each video. We recruited 30 participants and collected 3,657 interactions, each with temporal EEG signals during viewing as well as user behavior and multidimensional labels. Finally, we collected three types of data: user behavior log,  EEG signals, and self-assessment of six MAES. 

Subsequently, we present the statistical information of the dataset and show the rich information contained in the dataset. 
Besides, we discuss the possible applications for the dataset. We first show its impact on user understanding in the short video recommender system, with some primary discoveries thrown. 
We also establish benchmarks for rating prediction inferred from EEG signals and prevalent recommendation algorithms. Experiments show significant performance improvement with the inclusion of EEG signals, demonstrating the importance of introducing brain signals to recommender systems. 

These are our main contributions: 
\begin{itemize}
\item We proposed the first dataset that contains EEG signals in a real scenario of watching short video streaming. On the basis of user behavior, we provided multidimensional affective engagement scores~(MAES), which are valence, arousal, immersion, interest, visual and auditory, as explicit feedback. 
\item 
We establish benchmarks for rating prediction by the recommendation algorithm. Comparative experiments show significant performance improvement with the inclusion of EEG signals, demonstrating the importance of introducing brain signals to recommender systems. 
\item 
We show the perspective of understanding the affective experience and cognition activity behind user behaviors in the recommender system.

\end{itemize}


The remainder of this paper is organized as follows:
we review related datasets in Section~\ref{related}. 
Then we introduce our dataset and its collecting procedure in Section~\ref{dataset}.
Section~\ref{statistics} presents the combination and the statistical analysis for our dataset. Next, we conducted experiments to show the potential applications in Section~\ref{application}. 
Finally, Section~\ref{discussion} and Section~\ref{conclusion} discuss and conclude our work.

\section{Related Datasets}



\label{related}
In this section, we review the work of datasets in the short-video recommendation scenario and EEG datasets in affective computing and compare our dataset with theirs~(Table~\ref{tab:compare}). 

\begin{table*} [htbp]\small
\renewcommand\arraystretch{1.5}
\caption{Comparison of the EEG-SVRec with other datasets in the video/music recommendation and the video affective computing domain. \textbf{U\&I} represents user and item id. \textbf{Peri.Bio} represents peripheral biosignal~(such as, heartbeat, eye tracking, ECG). }
\centering
\setlength{\abovecaptionskip}{0.5cm}
\setlength{\belowcaptionskip}{0.5cm}
\begin{tabular}{@{}c|c|c|cccccc@{}}
\hline
\textbf{Domain} &  \textbf{Datasets}  &\textbf{Item/Stimulus} & \makebox[0.04\textwidth][c]{\textbf{U\&I}} & \makebox[0.09\textwidth][c]{\textbf{\small Impression}} & \makebox[0.07\textwidth][c]{\textbf{Ratings}} & \makebox[0.07\textwidth][c]{\textbf{Emotion}} & \makebox[0.07\textwidth][c]{\textbf{Peri. Bio}} & \makebox[0.05\textwidth][c]{\textbf{EEG}}  \\
 \hline
\multirow{7}{*}{\makecell{Recommendation\\(open domain)}} & Movielens &  Movie & \checkmark &  &  \checkmark &  &   \\
 & Toffee & Short Video & \checkmark & \checkmark  \\
& KuaiRand  & Short Video & \checkmark & \checkmark \\
 & MMSSL  & Short Video & \checkmark & \checkmark \\
 & Tenrec  & News, Short Video & \checkmark & \checkmark \\
 & Last.fm & Music & \checkmark & \checkmark & \\
 & MUMR  & Music & \checkmark & \checkmark & \checkmark & \checkmark & \checkmark &  \\
\hline
 \multirow{3}{*}{\makecell{Affective\\computing\\(closed domain)}} & DEAP &  1min music videos & \checkmark &  & \checkmark & \checkmark & \checkmark & \checkmark\\
& SEED  & 4min movie clips & \checkmark &  & \checkmark & \checkmark & \checkmark & \checkmark\\
& AMIGOS  & short and long movies & \checkmark &  & \checkmark & \checkmark & \checkmark & \checkmark\\
\hline
\makecell{Recommendation\\(open domain)} & \makecell{\textbf{EEG-SVRec}\\(ours)} & Short Video & \checkmark & \checkmark&\checkmark& \checkmark & \checkmark & \checkmark\\
 \hline
\end{tabular}
\label{tab:compare}
\end{table*}

\subsection{Dataset in Short Video Recommendation}

Short videos, a new type of online streaming media, have attracted increasing attention and have been one of the most popular internet applications in recent years. Thus, research on short video recommendations has gained traction, and some related datasets have been released. 

The datasets in short video recommendations are usually collected from online platforms with user id, item id, and their interaction behavior. 
An unbiased sequential recommendation dataset KuaiRand~\cite{gao2022kuairand} contains millions of intervened interactions on randomly exposed videos. 
Tenrec~\cite{yuan2022tenrec} is a large-scale and multipurpose real-world dataset with the item either a news article or a video. 
MicroLens~\cite{ni2023content} consists of one billion user-item interactions with raw modality information about videos. 
MMSSL~\cite{wei2023multi} is collected from the TikTok platform to log the viewed short videos of users. The multi-modal characteristics are visual, acoustic, and title textual features of videos. 
Some researchers conduct experiments on the Micro-Video dataset to validate their model~\cite{liu2021concept}. They construct a dataset by randomly sampling 100K users and their watched micro-videos over a period of two days. 
Other researchers crawled micro-videos from Jan 2017 to Jun 2018 from Toffee, a large-scale Chinese micro-video sharing platform~\cite{wei2019mmgcn}. 
Though the item is not a short video, the dataset Movielens ~\cite{harper2015movielens} interacting with the movie contains user ratings~(ranged 1-5), which have large scale and have had a substantial impact on education, research, and industry. 

Different from above, music dataset Last.fm-1k\footnote{https://www.last.fm/} represents the whole listening habits for nearly 1,000 users. MUMR~\cite{li2022towards} used a dataset in the music recommendation scenario with the collection of the contexts from low-cost smart bracelets. 
\citet{he2023understanding} consider immersion in online short videos with psychological labels, video features, and EEG signals. In contrast to them, we provide the dataset with various multidimensional affective engagements, giving a deep understanding of users. 

Since we collected from user studies, our data contains detailed video and audio features, behavior logs, user multidimensional affection engagement scores, and EEG and ECG signals.



\subsection{EEG Dataset in Affective Computing}

EEG~(Electroencephalogram) has been popular in neuroscience and psychology since it is a non-invasive technique used to measure the electrical activity of the brain. 
Utilizing physiological signals to help understand people's affection and cognition has become widespread in affection computing for its good balance between mechanistic exploration and real-world practical application. 
By analyzing EEG signals, researchers can identify patterns that are associated with different emotional states. 
Researchers collected EEG and peripheral physiological signals when using music, images, and videos as stimulation. Affection is annotated by the participants.

MIIR~\cite{stober2015towards} record the EEG signals from 10 participants when listening to and imagining~(by tapping the beat) 12 short music fragments. Then they rate their taping ability and familiarity. 
Images can also be the stimulus. A dataset in neuromarketing containing EEG signals of 14 electrodes from 25 participants and their likes/dislikes on e-commerce products over 14 categories with 3 images each~\cite{yadava2017analysis}.

The stimulation of videos includes both visual and auditory aspects, making the information more diverse and rich.
DEAP~\cite{koelstra2011deap} is the dataset of 32 participants whose EEG and peripheral physiological signals were recorded as each watched 40 one-minute excerpts of music videos. 
The SEED database~\cite{zheng2015investigating} contains EEG data of 15 subjects, which are collected via 62 EEG electrodes from the participants when they are watching 15 Chinese film clips with three types of emotions, i.e., negative, positive, and neutral.
Moreover, AMIGOS~\cite{miranda2018amigos} collected EEG, ECG, and GSR from 40 participants when watching 16 short videos and 4 long videos. Participants annotate their emotions during watching these videos with self-assessment of valence, arousal, control, familiarity, liking, and basic emotions. 

Datasets play a very important role in EEG affective computing. New methods and models have been proposed based on existing datasets to facilitate evaluation. 
However, the videos given to participants to watch are pre-selected for stimulating different emotions and are consistent between participants~(closed domain). Participants were unable to actively influence the video playback, e.g., slide done at any time to switch to the next video. 
Unlike them, our experiments take place in a real online short video browsing scenario, where videos come from among millions of videos on the platform and are presented to participants through personalized recommendation algorithms or in non-personalized or randomized ways~(open domain). During browsing, participants actively engage in behaviors such as swiping and liking videos. 

These existing research efforts show the application potential of EEG in various fields. 
In the context of short video recommendations, there still has no dataset to find the correlation between physiological signals and affective engagement during real short video scenarios. 
What we add on top of these works is that we conduct a user study where participants browse short videos in a real scenario and collect their behavior, multidimensional affective engagement labels, and EEG and ECG signals.

\section{Dataset Construction}
\label{dataset}

This section mainly covers ethical and privacy, participants, video stimuli material, apparatus, and experimental procedure~(browsing stage and labeling stage).


\subsection{Ethical and Privacy}

Our user study has underwent review and obtained approval from the institutional ethics committee, xxx University~(approve number: xxx~\footnote{The protocol ID is hidden for double-blinded review}). This study has undergone a rigorous ethical review process to ensure the protection of the participant's rights. In compliance with established ethical guidelines, we have taken multiple measures to protect the participants' privacy, including anonymizing the collected data and obtaining informed consent from all participants before the study. Furthermore, participants were fully informed about the study's objectives, procedures, and potential outcomes. The EEG data collection method employed in this research is non-invasive and poses no harm to the participants. This approach ensures that the study adheres to ethical standards while maintaining the integrity of the research findings. As for the item in the dataset, we only provide anonymized video ids, encoded video tags, and extracted video characteristics~(shown in Section~\ref{sec:cha_video}).

\begin{figure} 
    \centering
    \setlength{\abovecaptionskip}{0.1cm}
    \includegraphics[width=8.5cm]{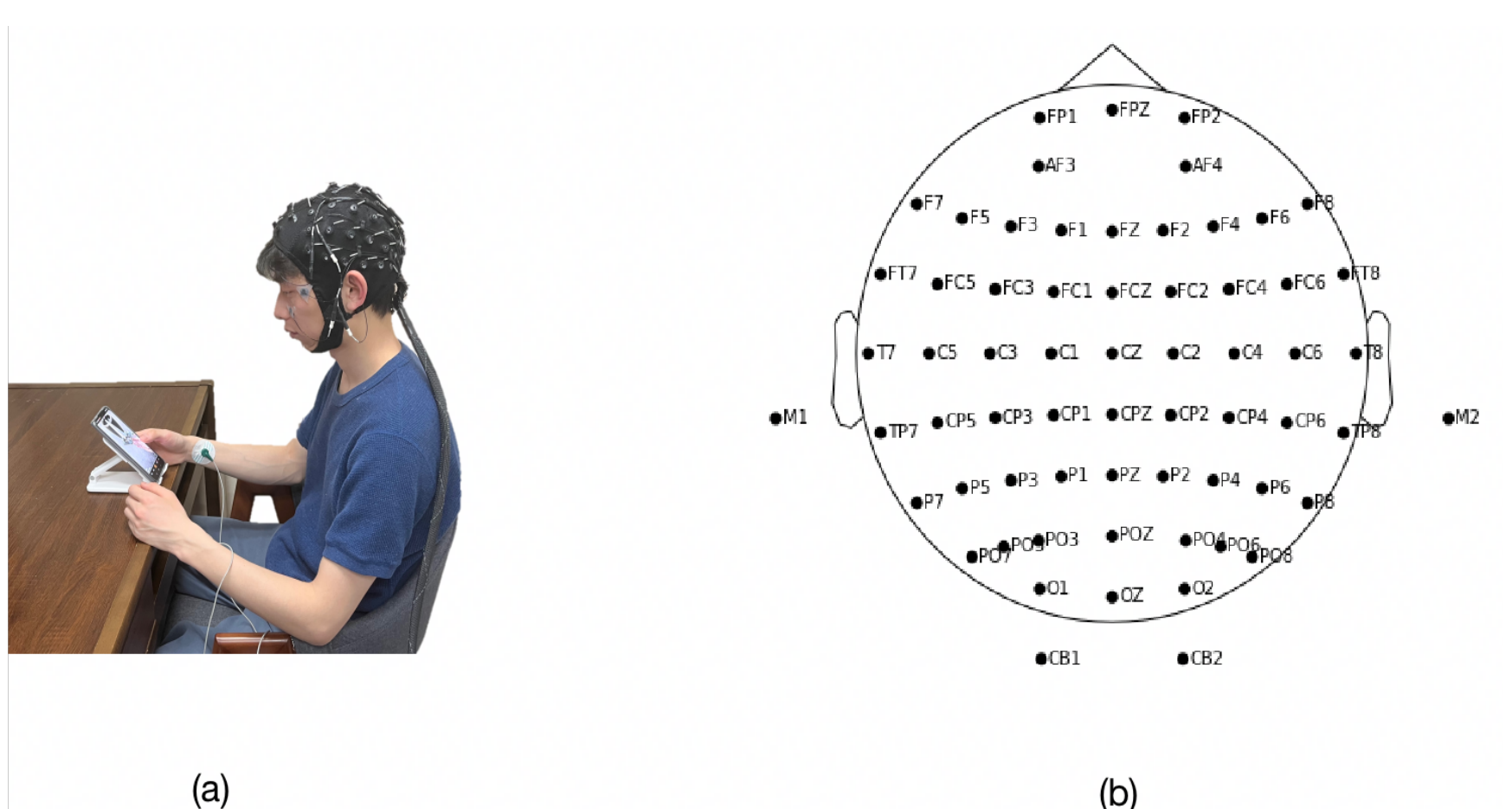}
\caption{EEG and ECG data acquisition setup: (a) A participant wears an EEG cap while watching short videos in a laboratory setting~(Image display has been approved).  (b) International 10-20 electrode placement standard for EEG. }


\label{fig:par_1020_ECG}
\end{figure}

\subsection{Participants}

\begin{figure*}[htbp]
    \centering
    
\setlength{\abovecaptionskip}{0.cm}
    \includegraphics[width=18cm]{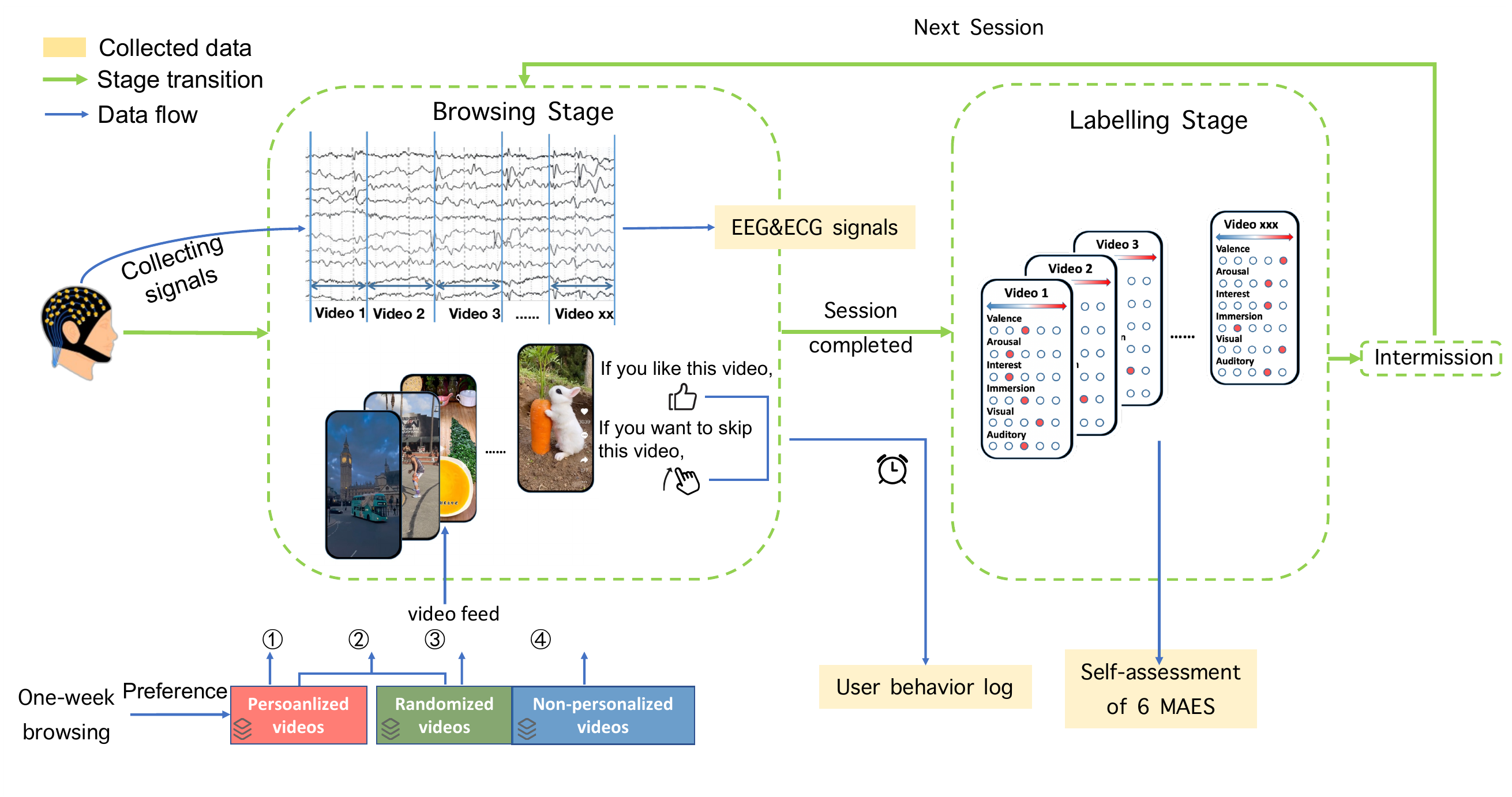}
\caption{The overall procedure of the lab study for data collection.}
    \label{fig:procedure}
\vspace{-0.5cm}
\end{figure*}

We recruited 30 college students aged between 18 and 30~(M=22.17, SD=2.20) for our study. The participant group consisted of 16 males and 14 females, majoring in various fields such as computer science, law, medicine, and sociology. All participants were familiar with at least one short video platform and used it at least once a day. To protect participants' privacy, we provided each participant with a new account on the short video platform. Each participant was required to participate in two experimental settings: a 10-hour preference collection phase and a 3-hour lab study phase as Figure~\ref{fig:par_1020_ECG}(a).~(including preparation and rest time). Upon completion of the experiments, each participant received approximately 60 dollars in research compensation.




\subsection{Video Stimuli Material}
\label{sec:material}

Participants browse short videos on a popular video platform, and all items are on the platform. 
The platform has two settings: personalized and non-personalized. Since they are all affected by the strategy of the platform, we present randomized videos as well. Thus, we categorized the short video stimuli to be presented to the participants into three video pools: personalized, non-personalized, and randomized.

The \textbf{personalized video pool} mainly consists of videos selected based on the preference information collected during the 10-hour preference acquisition phase for each participant, obtained through the short video platform's algorithm. 
The \textbf{non-personalized video pool}, with personalized-off, disregards user interaction history and distributes videos may be based on their current popularity ranking. It is worth mentioning that the videos in personalized and randomized pools have a duration of 30-60 seconds, while the non-personalized video pool's time restriction of 30-60 seconds was removed due to the distribution mechanism by platforms. 
The \textbf{randomized video pool} is sampled from the large video platform's video collection, filtered by different popularity levels. We first divided the large video pool into three levels based on view counts, and then randomly selected 100 videos from each level. After that, to ensure the category richness and healthiness of the selected short videos, we filtered 25 videos in each group. 

The selection of videos from these three pools results in different session compositions. 
Four distinct session modes were established: \textbf{personalized mode}, \textbf{randomized mode}, \textbf{mixed mode}, and \textbf{non-personalized mode}. 
It is clear that the personalized and randomized modes consist of 20-30 specific videos from their respective video pools, with a duration of 30-60 seconds each. 
In the mixed mode, an assortment of 20-30 videos is presented, with an equal proportion of personalized and randomized videos, maintaining a 1:1 ratio. Video sequences are random, ensuring a well-distributed and varied exposure for the study participants. 
The non-personalized mode involves extracting a certain number of videos from the non-personalized video pool. 

\subsection{Apparatus}



We used a smartphone with a 6.67-inch screen and a 120Hz refresh rate, which connected to a stable local area network (LAN) Wi-Fi to ensure network stability. Participants were allowed to adjust the screen brightness and device volume to a comfortable level before the experiment. They can also adjust the seat position and the angle of the smartphone to a suitable position. During the browsing stage, participants were required to minimize body and head movements to ensure the high quality of the collected physiological signals in Figure~\ref{fig:par_1020_ECG}(a). 
A Scan NuAmps Express system~(Compumedics Ltd., VIC, Australia) along with a 64-channel Quik-Cap~(Compumedical NeuroScan) was utilized for recording the participants' EEG data in Figure~\ref{fig:par_1020_ECG}(b)~\cite{homan1987cerebral}. Some electrode points were also used to eliminate head movement and other artifacts. The impedance of the EEG channels was calibrated to be under 10 k$\Omega$ in the preparation step, and the sampling rate was set at 1,000 Hz.


\subsection{Experiemental Procedure}

Each participant underwent 10-hour  preference information collection phase in a week, followed by laboratory experiment phase that included browsing and labelling stages. In the laboratory experiment, participants viewed 4 to 5 sessions of short videos, with each session comprising a 15-minute browsing stage and a roughly 10-minute labelling stage. After completing the video labelling for each session, participants were given a 5-minute rest before proceeding to the next session's browsing stage. 

During the \textbf{browsing stage}, participants watched sessions of short video sequences distributed from different video pools with each session comprised of 20-30 short videos. Throughout the short video browsing process, participants were allowed to interact with the videos primarily through \textbf{liking} and \textbf{swiping away} (the video). If participants enjoyed the video they were currently watching, they could click the \textit{like button} at any time during playback. Additionally, if participants did not wish to continue watching the video, they were allowed to \textit{swipe away} anytime. It's noted that the video will be replayed when done without swiping away. Electroencephalogram~(EEG) and electrocardiogram~(ECG) physiological signals were continuously collected. 

After each participant has completed browsing a short video sequence within a specific session, we conducted a video-level multidimensional affective engagement self-assessment \textbf{labelling stage}. Participants were given a brief recap of each video chronologically based on their browsing history. Subsequently, they rated each short video on a 5-point Likert scale across six multidimensional affective engagement indicators. the labelling instructions are given to the participants
\textsuperscript{\ref{link}}.
The six dimensions are valence, arousal, immersion, interest, visual, and auditory. 
\textit{Valence} represented the positive and negative aspects of emotions, while \textit{Arousal} indicated the intensity of emotions. 
\textit{Immersion} denoted the degree of the participant's involvement while watching the video, and \textit{Interest} indicated the extent to which the video aligned with the participant's personal interests. \textit{Visual} and \textit{Auditory} scores described the presentation quality of visual elements~(e.g., scenery, graphics) and auditory elements,~(e.g., voices, music). 

Our experiment collected MAES through questionnaires, gathering ratings for videos within each session after its completion. Participants were asked to recall the videos by viewing the first few seconds and to rate them across the six dimensions until they could adequately recall the video. In post-experiment interviews, participants reported that the number of videos per session did not cause memory difficulties, so they could generally recall the browsing history and complete labelling after watching the initial seconds of each video. Having completed the video labelling for each session, participants were given a 5-minute rest before proceeding to the browsing stage in the next session. Thus, the labelling stage generated a corresponding score for each of the six MAES for every short video.

Ultimately, we obtained three types of video-level data: browsing behavior logs, EEG and ECG signals, and multidimensional affective engagement self-assessment labelling. 

\section{Dataset Description}
\label{statistics}

In this EEG dataset, we ultimately collected 3,657 interactions from 30 users involving 2,636 items~(short videos). Due to the different participants watching the same short video in randomized mode, multiple interactions can be associated with the same item. Each interaction~(U-I pair) corresponds to a related EEG and ECG segment. Additionally, each interaction is associated with a behavioral log and a self-assessment of MAES.
To further describe the dataset, we introduce it from four aspects: the EEG signals, behavioral, self-assessment data, and characteristics of short videos.

\begin{table}[h]
\renewcommand\arraystretch{1.5}
\centering
\setlength{\abovecaptionskip}{0.5cm}
\setlength{\belowcaptionskip}{0.5cm}
\caption{The Statistics of Dataset. Each interaction has corresponding MAEs and EEG signals.}
\begin{tabular}{@{}c|ccccc@{}}
\hline
\textbf{}& \textbf{\#User} 
& \textbf{\#Item} & \textbf{\#Interaction} & \textbf{\#EEG datasize} \\
 \hline
EEG-SVRec & 30 & 2,636 & 3,657 & 62GB \\
 \hline
\end{tabular}
\label{tab:data_count}
\end{table}

\subsection{EEG statistics and preprocessing}
\label{sec:EEG_pre}

Here, EEG data are collected through all 3,657 interactions. For each interaction, the size of EEG data is (${Ch}$, ${fs}$ $\cdot$ ${T}$), where ${fs}$ is the sample rate~(1000 Hz), ${T}$ denotes the recording duration of the interaction, and ${Ch}$ is the number of electrode channels~(62 in total). We preprocess EEG data extract features as follows:









The raw EEG data is subjected to a series of preprocessing steps to eliminate noise and artifacts and enhance the signal quality. The preprocessing pipeline comprises the following stages. First, baseline correction: We first perform baseline correction to remove any constant offsets or drifts in the EEG signals, ensuring that the baseline amplitude is zero. Second, rereferencing: Re-referencing employs the average of M1 and M2 mastoid electrodes as the new reference, minimizing potential bias and improving the signal-to-noise ratio. Third, filtering: Filtering applies a 0.5 Hz to 50 Hz band-pass filter to remove low-frequency drifts~(<0.5~Hz) and high-frequency noise~(>50~Hz), as well as 50~Hz powerline interference. Last, artifact removal: Artifact removal eliminates abnormal amplitude signals and artifacts induced by eye blinks or head movements.

After the preprocessing steps, we proceed to extract features from the cleaned EEG signals. In this study, we focus on the extraction of differential entropy~(DE) as a feature, which has been shown to be useful in characterizing the complexity and information content of EEG signals~\cite{duan2013differential}.
Firstly, we estimate power spectral density~(noted as P(f)) using Welch’s method~\cite{welch1967use}~(sampling frequency is 1000) based on sliding window. The window length is two divided by the lower bound of the frequency band. 
Secondly, we normalized for each band and calculated DE using the following formula:

\begin{equation}
DE = -\int P(f) \log(P(f)) \, df
\end{equation}

The frequency bands are delta (0.5-4 Hz), theta (4-8 Hz), alpha (8-13 Hz), beta (13-30 Hz), and gamma (25-50 Hz). 
Finally, for each second of EEG signals, we extract a DE of each electrode and each frequency band. 





\subsection{User Behavior log and self-assessment of MAES}
\begin{figure}[htbp]
    \centering
    \setlength{\abovecaptionskip}{0cm}
    \includegraphics[width=8.5cm]{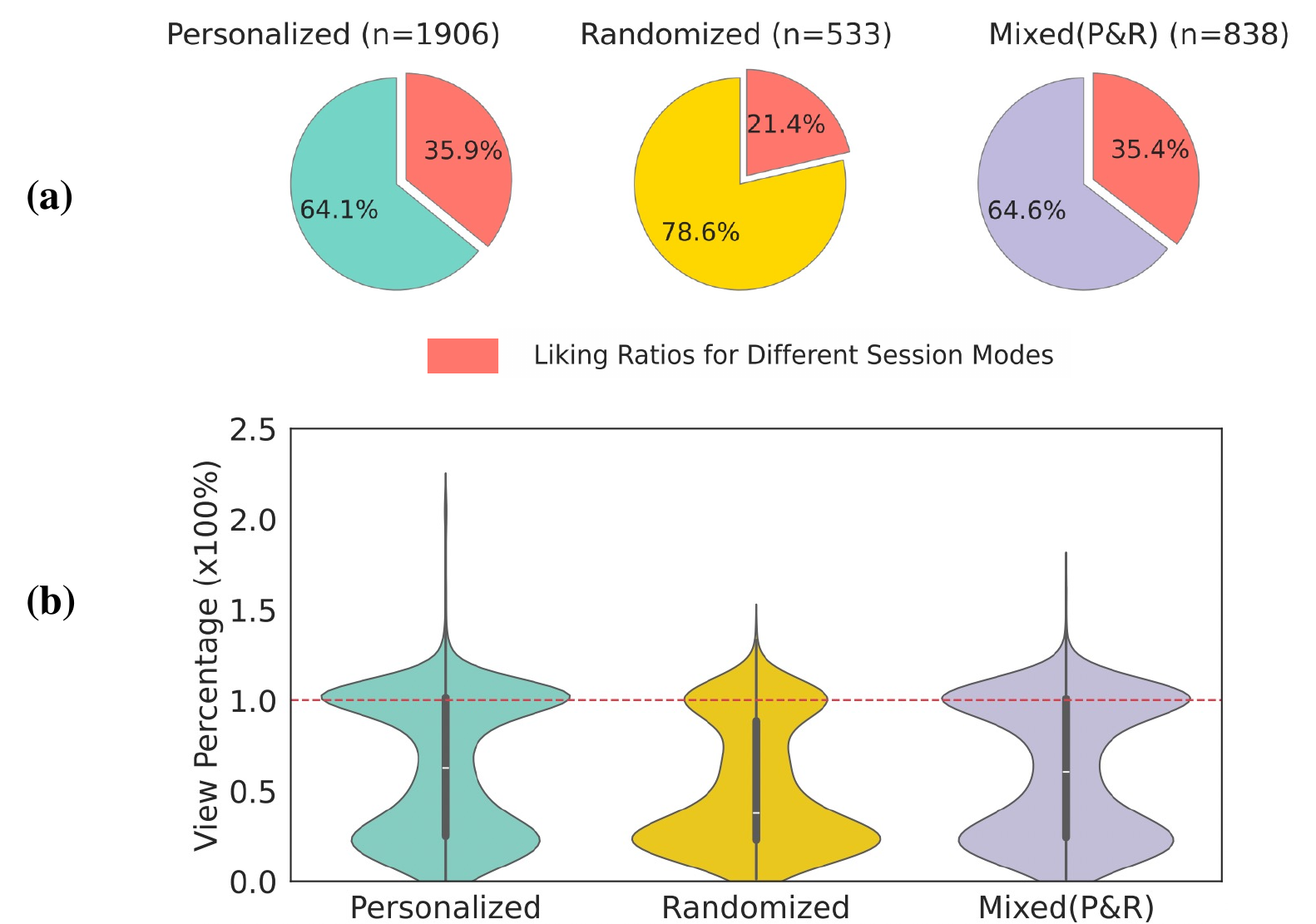}
\caption{(a) Proportion of likes for short videos: overall and across three session modes (personalized, randomized, and mixed). (b) View percentage distribution across different session modes~(View percentage is the viewing duration divided by the video duration. 1.0 represents viewing the video once.)}
\label{fig:behavior}
\end{figure}





After integrating the log and label files and corresponding them to the EEG via timestamps, we obtained each subject's interaction behavior~(liking and viewing duration) and MAES for their video viewing. 
For each interaction, the UNIX timestamps of browsing are aligned with the start and end time of the corresponding piece of the psychological signals. 
The video sequence and session mode are also important. Thus, we provide the order of the video in the interaction sequence and session mode~(Randomized, Personalized, and Mixed). As for the Mixed mode, we use further distinguish the personalized recommendation video from the random one.


In Figure~\ref{fig:behavior}~(a), we present the distribution of the proportion of likes for short videos in both the overall context and across three distinct session modes: personalized, randomized, and mixed (a combination of personalized and randomized). Notably, the like rate in the personalized mode~(35.9\%) and the mixed mode~(35.4\%) are relatively similar. In contrast, the like rate in the randomized mode~(21.4\%) falls below. Same as the like rate, view percentage in personalized mode and mixed mode in Figure~\ref{fig:behavior}~(b) is higher than randomized overall. It's surprising that the performance of mixed mode is relatively similar to the personalized mode. Likes and view percentages are presumably influenced by contexts in the session. Focusing on user experience from behavior may shed a little light on recommender systems. 


\begin{figure}[htbp]
    \centering

    \setlength{\abovecaptionskip}{0.cm}
    \includegraphics[width=8.5cm]{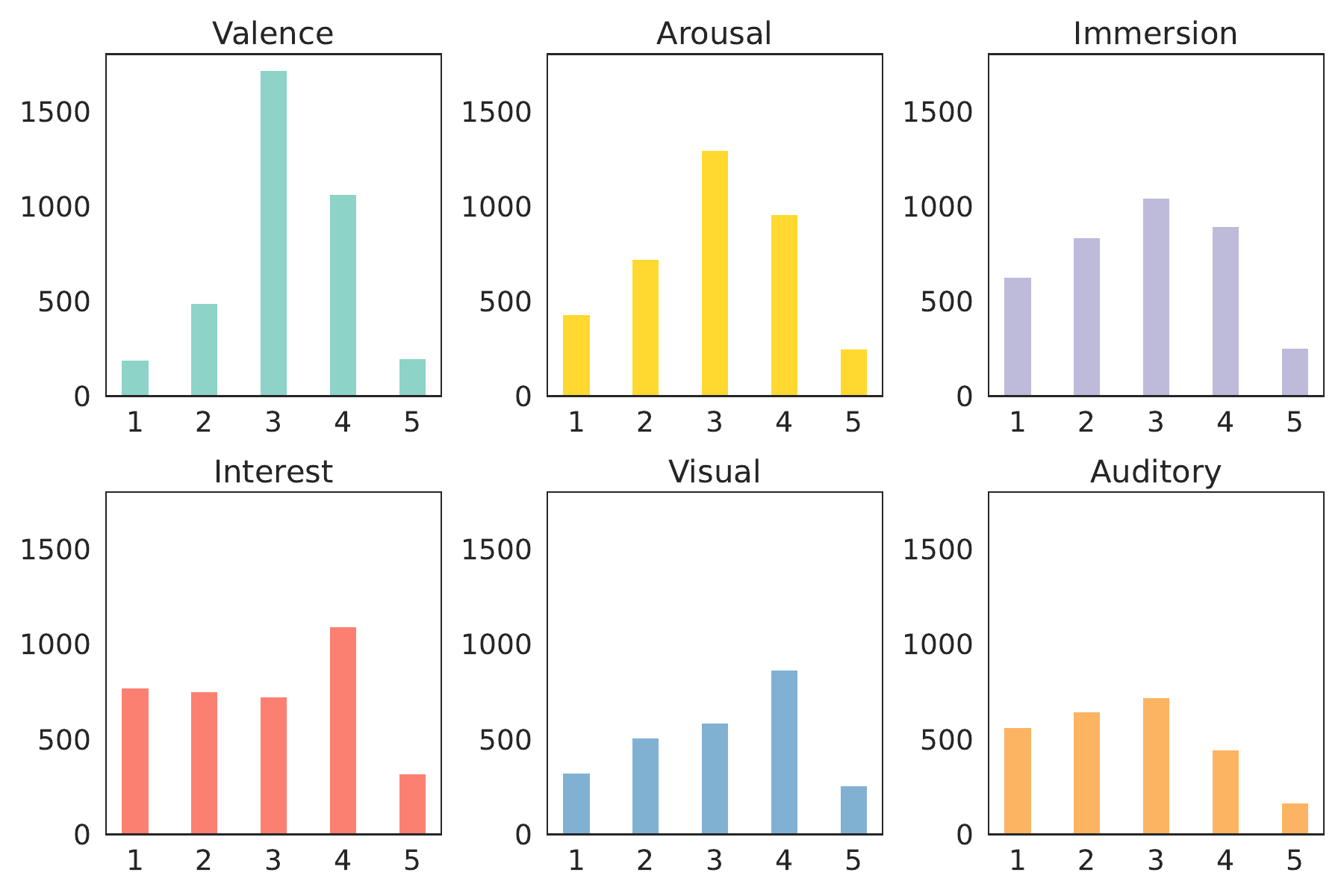}
\caption{The distribution for six MAES~(valence, arousal, immersion, interest, visual and auditory). }
\label{fig:dis_hot}
\end{figure}

In Figure~\ref{fig:dis_hot}, it can be observed that the distribution of the six MAES exhibits noticeable differences. 
It can be observed that valence and arousal, commonly utilized as two-dimensional indices in the field of emotion recognition, both exhibit a distinct distribution with 3 being the highest point. Immersion, interest, visual and auditory demonstrate a relatively uniform distribution compared to the former two, indicating a more effective differentiation in representing video content.




\subsection{Characteristic of Short Videos}
\label{sec:cha_video}
We meticulously extracted comprehensive video features, encompassing both visual and auditory aspects, to further investigate components related to audio and visual ratings.

For video featurization, we sampled each frame per second and computed an array of features. Specifically, we determined the mean (representing brightness) and standard deviation (representing contrast) by converting each frame to grayscale. Additionally, we assessed hue, saturation, value (in terms of HSV), Laplace variation, and color cast for each frame. Regarding audio, we initially extracted audio signals from the short videos using their native sampling rate. We then employed openSMILE to compute features from the ComParE2016 acoustic feature set~\cite{schuller2016interspeech}, maintaining the same sampling rate. Subsequently, we utilized the Audio Spectrogram Transformer~\cite{gong2021ast}, trained on AudioSet, to classify audio events with a sampling rate of 16,000 to comply with the classifier. If the event was classified as music, we employed Librosa to detect beats and determine the tempo.

\section{Example applications}
\label{application}


\subsection{Impact on User Understanding in Recommendation}

\begin{figure}[htbp]
    \centering
    \setlength{\abovecaptionskip}{0.cm}
    \includegraphics[width=7.5cm]{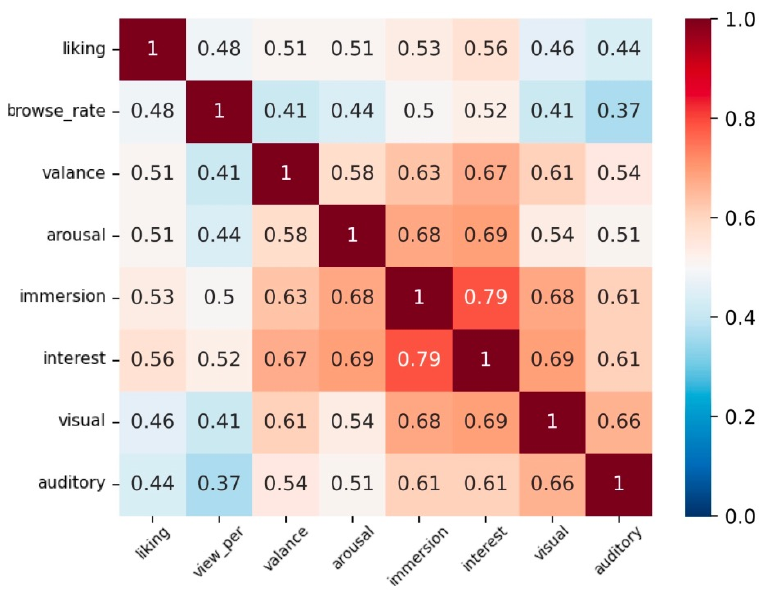}
\caption{ Heatmap presents the correlations of behavior~(liking, and view percentage) and MAES~(valence, arousal, immersion, interest, visual, and auditory). }
\label{fig:rating_corr}
\end{figure}

\subsubsection{Analysis of MAES and Browsing Behavior}

\begin{figure*}[htbp]
    \centering
    \setlength{\abovecaptionskip}{0.2cm}
    \includegraphics[width=15cm]{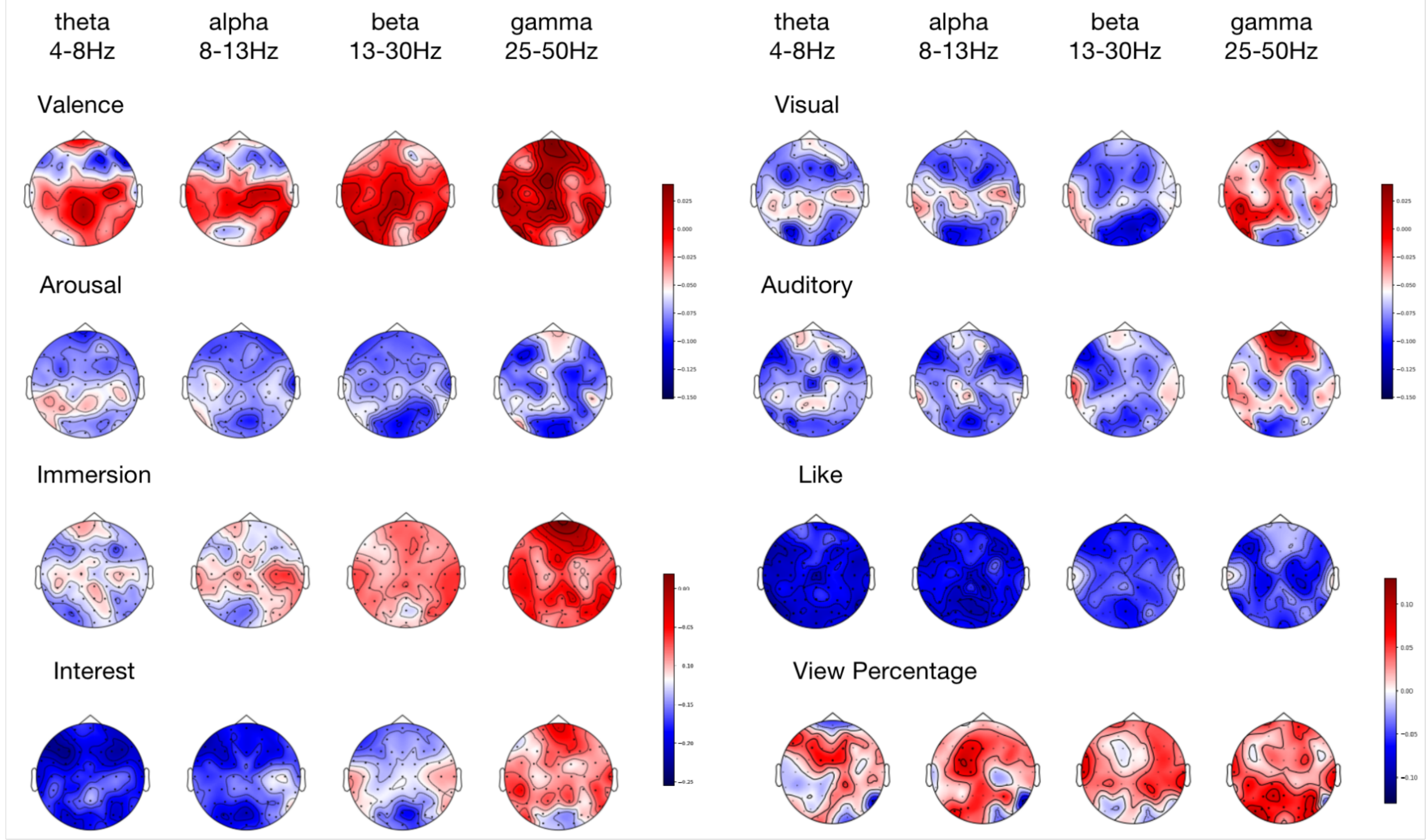}
\caption{The mean correlations (overall participants) of the MAES~(emotion of valence and arousal, immersion, interest, and rating of visual and auditory, ranged 1-5) and behaviors~(like and view percentage) with DE in the broad frequency bands of theta (4-8 Hz), alpha (8-12 Hz), beta (12-30 Hz), and gamma (30-45 Hz). The white circle marks the significant correlation~(p~\textless 0.05).} 
\label{fig:topo}
\end{figure*}

Figure~\ref{fig:rating_corr} presents the correlation between behavioral and MAES attributes. It can be observed that \textbf{Liking} has the strongest correlation with Interest~(0.56), followed by Immersion~(0.53) and Valence~(0.51). This suggests that the users' preferences are more closely related to their interest in the content and the degree of immersion they experience while viewing the video, rather than simply the valence or arousal induced by the content.
On the other hand, the View Percentage attribute exhibits the highest correlation with Immersion~(0.50) and Interest~(0.52), indicating that the percentage of browsing is more likely to be influenced by their interest in the content and the level of immersion they experience. This further highlights the importance of considering users' interests and immersion levels when designing recommender systems to improve user engagement and browsing experience.
The above findings emphasize the need to consider users' interests and the degree of immersion they experience when designing effective recommendation algorithms. We are expecting more findings to be discovered by the researchers. 

\subsubsection{The Relation of EEG with MAE and Behaviors}
Figure~\ref{fig:topo} displays the topographical maps illustrating the correlations between EEG signals and the six MAES as well as the two behaviors. These maps reveal distinct correlation patterns for each MAES and behavior. Furthermore, some unique findings emerge, such as the consistent presence of strong correlations between gamma-band electrodes in the frontal lobe area across all six emotion annotations.

Gamma waves are known to play a critical role in numerous brain functions and cognitive processes, including attention, memory, perception, and consciousness~\cite{li2009emotion, meador2002gamma}. The activation of gamma waves in the frontal lobe suggests the involvement of this region in the associated cognitive processes. As a key area of the brain, the frontal lobe is closely linked to higher cognitive functions, such as decision-making, planning, problem-solving, working memory, and attention control~\cite{fuster2002frontal}. The observed activation of gamma waves in the frontal lobe may be indicative of the engagement of these higher cognitive functions during the tasks.

\subsection{Recommendation in Terms of Various User Feedback Signals}
\begin{table*}[htbp]
\renewcommand\arraystretch{1.5}
\caption{The recommendation performance~(in terms of AUC) that leverages liking, interest, immersion, visual preference, and auditory preference as user feedback respectively. The two-sided t-test is conducted
. $^{*}$ indicates p-value < 0.05. \textbf{bold} shows the higher result of the two settings. }
\centering
\setlength{\abovecaptionskip}{0.5cm}
\setlength{\belowcaptionskip}{0.5cm}
\begin{tabular}{@{}cc|ccccccc@{}}
\hline
\textbf{Model}& \textbf{Feature} &  \textbf{Like} & \textbf{Immersion} & \textbf{Interest} & \textbf{Valence} & \textbf{Arousal} & \textbf{VisualPref} & \textbf{AudioPref}\\ 
 \hline
 FM & id  & 0.7152 & 0.6776 & \textbf{0.6950} & 0.6348 & 0.6917 & 0.6685 & 0.6419\\
 & id+EEG  & \textbf{0.7312}$^{*}$ & \textbf{0.6857} & 0.6933 &\textbf{0.6492} & \textbf{0.6929} & \textbf{0.6690} & \textbf{0.6675}$^{*}$\\
 \hline
DeepFM & id  & 0.7331 & 0.6869 & 0.7005 & 0.6379 & 0.6930 & 0.6691 & 0.6600\\
 & id+EEG  & \textbf{0.7368} & \textbf{0.6927} & \textbf{0.7010} & \textbf{0.6586}$^{*}$ &  \textbf{0.7077} & \textbf{0.6711} & \textbf{0.6608} \\
 \hline
AFM & id  & 0.7188 & 0.6774 & \textbf{0.6935} & 0.6406 & \textbf{0.6962} & \textbf{0.6736} & 0.6251\\
 & id+EEG  & \textbf{0.7236} & \textbf{0.6955}$^{*}$ & 0.6910  & \textbf{0.6583}$^{*}$ &  0.6898 & 0.6688 & \textbf{0.6578}$^{*}$\\
\hline
WideDeep & id  & 0.7324 & 0.7033 & 0.7027 & 0.6651 & 0.7066 & 0.6718 & 0.6735\\
  & id+EEG  & \textbf{0.7387} & \textbf{0.7056} & \textbf{0.7121} & \textbf{0.6660} & \textbf{0.7094} & \textbf{0.6978}$^{*}$ & \textbf{0.6767}\\
\hline
DCN-V2 & id  & \textbf{0.6937} & 0.6190 & 0.6698 & 0.5855 & 0.6340 & 0.6443 & \textbf{0.6585}\\
  & id+EEG  & 0.6924 & \textbf{0.6582}$^{*}$ & \textbf{0.6802}$^{*}$ & \textbf{0.6249}$^{*}$ & \textbf{0.6715}$^{*}$ & \textbf{0.6585} & 0.6440 \\
\hline
\end{tabular}
\label{tab:recommender}
\end{table*}

The proposed \textbf{EEG-SVRec} is also feasible for personalized recommendation tasks. Beyond the traditional way of taking \textit{liking} as the user feedback signal, various user feedback signals provided in the dataset can be leveraged as the ground truth.
We conduct experiments for 
item recommendation task while leveraging \textit{liking, immersion, interest, valence, arousal, visual preference}, and \textit{auditory preference} as user feedback signals, respectively. As an example, we 
provide the benchmark for item recommendation with and without EEG information. We use the popular recommendation toolkit Recbole~\cite{zhao2021recbole} for different algorithms, which only support point-wise evaluations for context-aware recommendation models rather than ranking-based evaluation, and report AUC scores but not NDCG performances. 

The dataset is split into training set, validation set, and test set by 7:1:2. 
As for the EEG data, we utilize the 310-dimensional~(62 channels * 5 frequency bands) DE~(EEG feature described in Section~\ref{sec:EEG_pre}) corresponding to interactions and project them into an embedding through a fully connected layer. 
We tune hyperparameters and choose the best result for each setting~(id and id+EEG). 

From the Table~\ref{tab:recommender}, it is observed that in most instances, models incorporating EEG signals achieve superior results, suggesting the general potential of EEG signals in recommendation tasks. 
It is worth noting that only simple way of introducing EEG information is implemented in the benchmark experiments, which directly embeds EEG signals as features, and has already  
effectively enhanced recommendation performance. This verifies that EEG contains additional valuable information
. Thus, leveraging EEG signals presumably assists recommender systems in better understanding user multidimensional affective engagement and behaviors, thereby providing 
better personalized recommendations.


EEG reflects the cognitive activity of viewing short videos, which can be used as auxiliary information to enhance representations. 
Thus, a natural idea is to enhance user and item embeddings with their corresponding EEG signals. 
The idea is widely used in existing recommendation models, such as review-based~\cite{chen2018neural, sun2020dual}, social-based~\cite{wu2019neural}, knowledge graph-based~\cite{chen2019co}, and visual-based~\cite{he2016vbpr} models. However, EEG directly reflects the user's brain activities, which can bring more in-depth user understanding 
beyond the above auxiliary information.
This opens a novel avenue
to enrich the representation of items and further help the recommender systems understand the users with 
incognizable, subject, and direct feedback with cognitive information. 
The comparison of EEG data and other information, as well as more sophisticated recommendation models, are left as future work. 

\section{Discussions and Limitations}

\label{discussion}
\subsection{Possible Research Directions}
In this section, we discuss the potential applications of the EEG-SVRec dataset in various aspects of short video recommendation systems and beyond.

(1) \textbf{Human-centric Evaluation Metrics}: The dataset offers a more human-centric perspective on evaluation metrics, going beyond traditional measures such as dwell time and likes. It enables researchers to assess recommender systems based on their ability to enhance users' overall experience, considering multidimensional aspects of user engagement, rather than merely maximizing utility metrics.

(2) \textbf{Uncovering the Relationship Between User Behavior and Cognitive Acitvity}: Utilizing the dataset to study user behavior and cognitive activities during the recommendation process can reveal insights into how brain activity can inform adjustments in recommendations. This knowledge potentially helps reduce information echo chambers and enhance content diversity, leading to a more balanced and varied user experience. 

(3) \textbf{EEG-guided Recommendation Algorithms}: The EEG-SVRec dataset opens up opportunities to explore the development of EEG-guided recommendation algorithms that incorporate EEG signals for a deeper understanding of user preferences and behavior. By leveraging a smaller labeled EEG dataset alongside a larger unlabeled dataset, algorithms can potentially learn more accurate and personalized recommendations by generalizing the knowledge gained from EEG signals across a broader user base. Furthermore, EEG reflects the cognitive activity of viewing short videos which can be used as auxiliary information to enhance representation. 

(4) \textbf{Accessibility for Users with Disabilities in Short Video Streaming}: The EEG-SVRec dataset has the potential to facilitate the development of more inclusive recommendation systems tailored for individuals with disabilities. By analyzing the unique cognitive and emotional experiences of these users through EEG data, algorithms can be adapted to better cater to their needs and preferences, ultimately improving their experience with short video recommendations.

In summary, the EEG-SVRec dataset presents an array of potential applications that can contribute to the development of more effective, personalized, and inclusive recommendation algorithms. By focusing on a more human-centric approach and leveraging the rich information provided by EEG signals, researchers and practitioners can drive innovation in the field of recommender systems and enhance user experiences across various contexts.


\subsection{Limitations}

In this study, we present the EEG-SVRec dataset. Despite its potential value for the recommender systems community, there are some limitations that should be considered: 

(1) \textbf{Sample Size:} The dataset was constructed with a scale of 30 participants from the university, which may not fully capture the diversity of users on social media platforms. Although the sample size might seem limited, it is important to note that the high cost associated with EEG data collection can hinder the ability to gather larger sample sizes. Many published EEG datasets are with the same scale of participants universities~\cite{koelstra2011deap, savran2006emotion, zheng2015investigating}. 

(2) \textbf{Generalizability:} 
EEG's applicability in large-scale real-world scenarios could be challenging due to the required equipment and expertise. 
Meanwhile, personalized and randomized videos are 30-60s, which may differ from general contexts. The reason to choose 30-60s refers to Section~\ref{sec:material}. Despite this, investigating the temporal dynamics of user behavior and emotions in various recommendation settings would be a valuable direction for future research. 

(3) \textbf{Algorithmic bias:} The EEG-SVRec dataset might contain biases from the underlying recommendation algorithms from the platform, which could impact the generalizability of the findings. However, we provide the interaction with randomized video as unbiased data for this purpose. It is essential for future research to identify and address any potential biases present in the dataset. 




Despite these limitations, the EEG-SVRec dataset provides a valuable resource for exploring user behavior and emotions in short video recommendations and can inspire further research in this area.


\section{Conclusion and Future Work}

\label{conclusion}
This paper introduces EEG-SVRec, a novel dataset including EEG and ECG signals, multidimensional affective engagement annotations, and user behavior data for short video recommendation. This dataset bridges a critical gap by providing insights into user intrinsic experience and behavior in real-world short video scenarios. Our key contributions include proposing the first EEG dataset in short video streaming scenario, collecting multidimensional affective engagement scores, and providing both implicit and explicit user feedback. We carried out a rigorous experimental process for 30 participants and obtained a dataset, which is highly versatile and applicable to various research problems. We establish benchmarks for rating prediction by including EEG signals and prevalent recommendation algorithms. Experimental results demonstrate the usefulness of EEG signals in recommendation scenarios. It is worth noting that our current application of EEG signals is primary, leaving room for future improvements.

For future work, it is expected that more sophisticated models, such as DGCNN~\cite{song2018eeg}, could be employed to utilize electrode position information from the EEG signals and further improve recommendation performance on the EEG-SVRec dataset. By leveraging more advanced techniques, deeper insights into the role that EEG signals play in short video recommendation systems could be uncovered. Furthermore, the application of EEG and ECG signals could be expanded to a broader range of research areas, such as developing more affective-centric evaluation metrics and applications for individuals with disabilities. Lastly, the dataset holds significant societal value in further exploring the occurrence and changes in user emotions and cognitive behavior within short video recommendation scenarios. We anticipate that our work will inspire further exploration and innovation in the field of recommendation and encourage researchers to delve into these potential applications.

\clearpage
\bibliographystyle{ACM-Reference-Format}
\bibliography{reference}

\end{document}